\documentclass[prd, aps, superscriptaddress, preprintnumbers, twocolumn, floatfix, nofootinbib]{revtex4-2} 
\pdfoutput=1

\usepackage{amsfonts}
\usepackage{amsmath}
\usepackage{amssymb}
\usepackage{bm}
\usepackage{dcolumn}
\usepackage{graphicx}
\usepackage[latin1]{inputenc}
\usepackage{latexsym}
\usepackage{rotating}
\usepackage{hyperref}
\usepackage{graphicx}
\usepackage{color}

\newcommand\be{\begin{equation}}
\newcommand\bea{\begin{eqnarray}}
\newcommand\ee{\end{equation}}
\newcommand\eea{\end{eqnarray}}

\def\doi{http://doi.org}

\begin{document}

\title{On the Origin of the $SO(9) \rightarrow SO(3) \times SO(6)$ Symmetry Breaking in the IKKT Matrix Model}

\author{Robert Brandenberger}
\email{rhb@physics.mcgill.ca}
\affiliation{Department of Physics, McGill University, Montr\'{e}al, QC, H3A 2T8, Canada}
 
\author{Julia Pasiecznik}
\email{Julia.Pasiecznik@mail.mcgill.ca}
\affiliation{Department of Physics, McGill University, Montr\'{e}al, QC, H3A 2T8, Canada}

\date{\today}

\begin{abstract}
 
We suggest a dynamical mechanism which explains why in the supersymmetric IKKT matrix model the $SO(9)$ symmetry of the Lagrangian is spontaneously broken to $SO(3) \times SO(6)$, allowing only three large classical spatial dimensions to emerge.  The argument relies on the identification of D1-strings as stable excitations of the matrices about the cosmological background, and the absence of forces between parallel D1-string excitations. A spatial dimension can only become large if the D1-strings winding that direction can annihilate.  In the absence of forces, this requires the string world sheets to intersect,  a process which is highly suppressed if more than three dimensions of space are large. Implications for a recently proposed matrix cosmology scenario are discussed.
 
\end{abstract}

\maketitle

\section{Introduction}

Perturbative superstring theory is mathematically consistent only in a $1+9$-dimensional space-time. In this context, an urgent question is to explain the fact that we only see three large spatial dimensions.  Traditionally, one assumes that the six dimensions of space which we do not detect are compactified, and that there are mechanisms to prevent the dimensions of the compact space from becoming large. However, this assumption is not very satisfactory since we would like to understand the reason for the particular number of dimensions which are compact, and the origin of the separation of scales.

In \cite{BV}, a dynamical mechanism was proposed which can explain both the fact that exactly three dimensions of space become large,  and why the other dimensions remain compactified at a scale which typically is of the order of the string scale \footnote{See also \cite{ACD, MS, Randall} for some other ideas.} The proposal of \cite{BV} is based on {\it{String Gas Cosmology}} (SGC), a toy model for early universe cosmology based on considering the effects of new degrees of freedom and new symmetries which distinguish string theories from point particle theories (see \cite{SGCRevs} for reviews, and \cite{Perlt} for some early work).

The idea of SGC is to consider a background space-time filled with a gas of closed strings in thermal equilibrium.  This is a generalization of the starting point of standard cosmology with point particles replaced by strings. For simplicity, it was assumed that all spatial dimensions are toroidal with radius $R$.  Strings have degrees of freedom which point particles do not have: string winding modes and a tower of string oscillatory modes. A consequence of the presence of the string oscillatory modes is that there is a maximal temperature of a gas of closed strings in thermal equilibrium, the Hagedorn temperature $T_H$ \cite{Hagedorn}. Because the spectrum of string states contains winding modes. the theory is invariant under the interchange of $R$ and $1/R$ (in string units). Since the energy of string momentum modes scales as $1/R$ while that of the string winding modes scales as $R$, the mass spectrum of perturbative string states is invariant under the above symmetry. String interactions also obey the symmetry, and assuming that the symmetry extends beyond perturbation theory leads to the prediction that the theory must contain D-brane states (see e.g. \cite{Pol} for a textbook discussion).  

In \cite{BV}, the evolution of the temperature $T$ of a gas of closed heterotic strings was studied as a function of $R$. The result is sketched in Fig. 1: at large values of $R$, the energy is in the momentum modes which are light in this limit. As $R$ decreases, the energy density rises until the point when the oscillatory modes can be excited. At that point, the increase of $T$ levels off at a value slightly lower than $T_H$ (how much lower depends on the total entropy of the system). Once $R$ decreases below $1$, the energy flows into the winding modes and $T$ decreases. yielding
\be
T(R) \, = \, T(1/R) \, .
\ee

\begin{figure}
\includegraphics[scale = 0.3]{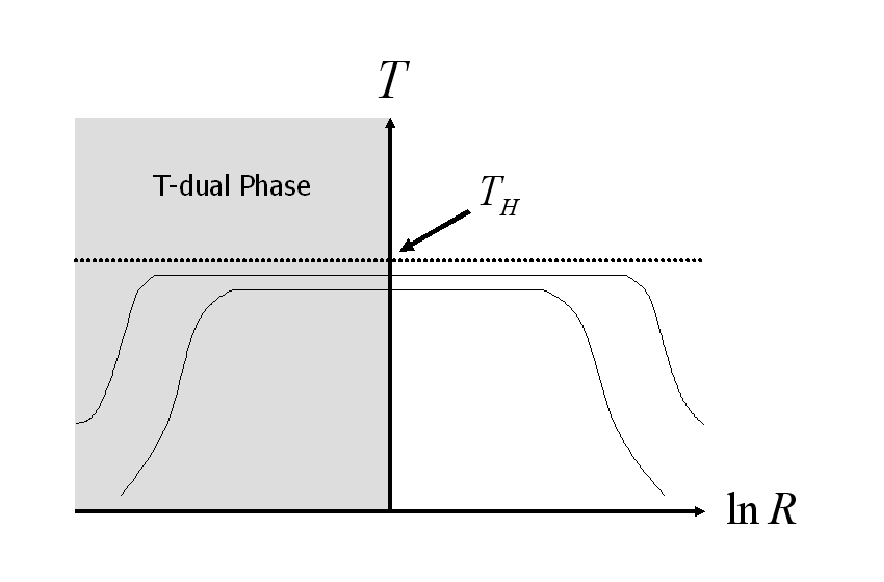}
\caption{Temperature $T$ (vertical axis) of a box of closed heterotic strings as a function of the box radius $R$(horizontal axis. $T_H$ is the Hagedorn temperature. Note that the range of values of $R$ over which the temperature loiters close to $T_H$ increases as the entropy of the system increases (figure from \cite{Jiro}).} 
\label{matrix}
\end{figure}

Based on these thermodynamic considerations, it was suggested in \cite{BV} to consider an initial thermal state of strings close to the Hagedorn temperature.  Based on Fig. 1 it is then reasonable to suppose that the universe will loiter for a substantial amount of time in this Hagedorn phase, yielding a time evolution of the cosmological scale factor as depicted in Fig. 2.  This is an {\it{emergent Universe}} scenario \footnote{Note that the same term has been introduced in a different context in \cite{Ellis}.}.  

\begin{figure}
\includegraphics[scale = 0.5]{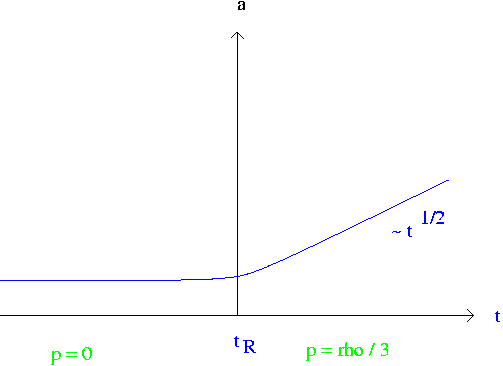}
\caption{Sketch of the evolution of the cosmological scale factor of the three large spatial dimensions in String Gas Cosmology.  The time $t_R$ is when the winding modes about these three dimensions annihilate.} 
\label{matrix}
\end{figure}

If the spatial sections are toroidal \footnote{Generalizations of the spatial topology were considered in \cite{Greene}. }, then the gas of strings in the Hagedorn phase will contain strings winding each dimension. Free energy considerations then indicate that spatial dimensions can only expand if it is possible for the string winding modes to decay. As discussed in \cite{BV}, if there are no long range forces between winding strings, then string world sheets will have to intersect in order for winding modes to be able to decay into string loops. The probability for this to happen vanishes if there are more than three large (compared to the string length) dimensions of space.  This yields a dynamical mechanism to understand the fact that we see three spatial dimensions to be macroscopic. Numerical evidence for this scenario was presented in \cite{MS2}, and isotropization of the three large dimensions was studied in \cite{Scott1} \footnote{In \cite{Moduli}, the stabilization of the moduli associated with the extra dimensions through the interaction of string winding and momentum modes was studied.  Assuming the mechanism of gaugino condensation \cite{gaugino} it has also been shown that the dilaton can be stabilized \cite{Frey}, while leading to supersymmetry breaking at the string scale \cite{Wei}.}.

As realized in \cite{NBV}, thermal fluctuations in the Hagedorn phase yield scale-invariant spectra of curvature fluctuations and gravitational waves. The spectrum of curvature fluctuations has a slight red tilt while that of gravitational wave has a characteristic blue tilt \cite{NBPV}. Hence, SGC has to potential to provide an alternative to inflationary cosmology for explaining the origin of cosmological structure (see e.g. \cite{RHBrev} for an extensive discussion of this point) with a prediction (the blue tilt of the gravitational wave spectrum) with which the scenario can be distinguished from canonical inflation models (inflation in the context of General Relativity coupled to matter obeying the Weak Energy Condition). Note that the resulting cosmology obeys the {\it Trans-Planckian Censorship Criterion} (TCC) \cite{TCC}, unlike standard effective field theory models of inflation.

While the mechanism for the origin of three large spatial dimensions from String Gas Cosmology is suggestive, the problem is that the dynamics for the Hagedorn phase was put in by hand.  In the context of Einstein gravity,  a thermal high density state could not be quasi-static, and adding a dilaton does not dramatically change the conclusions. Pre-Big-Bang Cosmology \cite{PBB} and Double Field Theory \cite{DFT} are approaches to consider the cosmology which result when taking into account all of the perturbative massless modes of string theory.  But no Einstein frame emergent scenario was obtained \cite{Gui}.  However, in the same way that at high densities point particle theories need to be described by their non-perturbative formulation, namely quantum field theory, we should expect that to describe a stringy early universe we need to start from a non-perturbative definition of superstring theory.

In the late 1990s a number of proposals for a non-perturbative formulation of superstring theory were made. In particular, the BFSS matrix model \cite{BFSS} was proposed as a non-perturbative definition of M-theory, and the IKKT matrix model \cite{IKKT} was proposed as a definition of Type IIB superstring theory.  The two matrix models can be related by dualities in the same way that M-theory and Type IIB string theory can be related by dualities in the low energy limit (see e.g. \cite{Wati, Ydri} for reviews of these matrix models, and \cite{Sumit} for the study of a lower dimensional model).  These matrix models involve nine ``spatial'' $N \times N$ Hermitean matrices $X_i$ and a temporal matrix $X_0$, also a Hermitean $N \times N$ matrix, plus their supersymmetric partners.The action has a $SO(9)$ symmetry under rotations of the spatial matrices.

A non-perturbative formulation of superstring theory should yield an emergent space-time with important implications for cosmology. In \cite{SB1, SB2} a proposal was made to obtain an emergent space-time starting from the BFSS matrix model in a high temperature state, as will be reviewed below.  At high temperatures, the Matsubara zero modes of the bosonic matrices dominate the action.  The action of the Matsubara zero modes of the BFSS matrix model is identical to the action of the bosonic sector of the IKKT matrix model.  Over the past twenty years there have been many numerical studies of the IKKT matrix model (see e.g. \cite{JNrev} for reviews).  The eigenvalues of the temporal matrix $X_0$ can be viewed as {\it emergent time}, while the matrices $X_i$ yield the nine spatial dimensions of Type IIB superstring theory.  There is evidence \cite{IKKTpt} of a $SO(9) \rightarrow SO(3) \times SO(6)$ phase transition after which the {\it extent of space parameters} in eactly three dimensions become large while the others remain small. The evidence comes from numerical simulations and is supported in part by Gaussian expansion computations of the free energy. Numerical studies, however, require the introduction of cutoffs, and the numerical results are to date not robust against changes of the numerical techniques.  It is hence desirable to obtain a physical understanding of the origin of the phase transition.

The goal of this paper is to discuss a possible mechanism which can explain the $SO(9) \rightarrow SO(3) \times SO(6)$ phase transition.  Our starting point is the realization that the dynamics of the phase transition is very similar to what is seen in SGC.  In SGC only three dimensions of space can become large since in higher dimensions string world sheets have measure zero probability of intersecting, and such intersections are required in order to get rid of string winding modes.  While string winding modes are present,  the corresponding spatial sections cannot expand since this would require energy. We postulate that it is the same mechanism which is at work in the IKKT matrix model. We identify D1-strings as solitonic excitations of the IKKT matrix model, as already proposed in \cite{IKKT} (see also \cite{Miao, Zarembo}). In the absence of a nontrivial background, it can be shown \cite{IKKT}, as will be reviewed in Section III, that as a consequence of supersymmetry there is no long range forces between parallel D1-strings. Free energy considerations tell us that in the presence of D1-strings winding a particular spatial dimension, this dimension cannot expand. Hence, as in SGC, the winding modes can only decay in at most three spatial dimensions, hence providing an explanation for the nature of the symmetry breaking phase transition.  In this paper, we will provide indications that this argument can be generalized to our cosmological background.

In the following section we review the IKKT matrix model and its connection with the emergent metric space-time construction of \cite{SB1, SB2} (see \cite{BBL3} for a short review).  The bulk of the analysis is in Section III. We first review how in the absence of a nontrivial background,  a set of two parallel D-strings can be embedded in the matrix model. We then propose a generalization of this construction in the case of a nontrivial background (specifically with the background of \cite{SB1, SB2} in mind). We compute the one loop effective action about this matrix configuration and show that, as a consequence of supersymmetry of the matrix model action, the result is independent of the distance between the D-strings \footnote{Note that the background is not supersymmetric.}. This demonstrates that, at least to leading order, the force between parallel D-strings vanishes.  In Section IV we show how the results of Section III lead to the $SO(9) \rightarrow SO(3) \times SO(6)$ symmetry breaking and to a dynamical decompactification of a three dimensional space.

Note that we are always working with the Minkowski signature model.

\section{Review of the IKKT Matrix Model}

The IKKT model \cite{IKKT} is a pure supersymmetric matrix theory given by the action
\be \label{IKKTaction}
S_{IKKT} \, = \,  -\frac{1}{g^2} {\rm{Tr}} \bigl( \frac{1}{4} [A^a, A^b] [A_a,A_b] + \frac{i}{2} \bar{\psi}_\alpha ({\cal{C}} \Gamma^a )_{\alpha\beta} [A_a,\psi_\beta] \bigr) \, ,
\ee
where $A^a$ ($a = 0, \cdots, 9$) are $N \times N$ Hermitean bosonic matrices,  $\psi_{\beta}$ ($\beta = 1, \cdots, 16$) are Hermitean fermionic matrices, the $\Gamma^a$ are the gamma matrices for $D = 10$, and $C$ is the charge conjugation matrix.  The vector matrix indices are raised and lowered with the Minkowski symbol $\eta_{\mu \nu}$, and $g$ is a coupling constant.  As was argued in \cite{IKKT}, in the limit $N \rightarrow \infty$ and with $\lambda \equiv g^2 N$ held fixed, this model yields a nonperturbative definition of Type IIB superstring theory. From the point of view of matrix theory, the choice of dimension $D = 10$ is distinguished in that it is only in this dimension that a normalizable zero energy state exists \cite{JF} \footnote{It would be interesting to explore the connection between this fact about supersymmetric matrix models and anomaly cancellation in superstring theory which leads to the number of space-time dimensions in perturbative string theory.}. Note that the action of this model has $SO(9)$ symmetry.

The partition function $Z$ of the Lorentzian IKKT model is given by the functional integral
\be
Z \, = \, \int {\cal{D}}A {\cal{D}}\psi  e^{i S_{IKKT}} \, ,
\ee
where the integration measures are the standard ones.  Based on this partition function, expectation values of matrix operators can be studied.  Since the matrices are Hermitean, it is possible to choose a gauge in which one of them is diagonal. Following \cite{Nish1} we pick the gauge in which $A^0$ is diagonal. The diagonal elements can be identified as instances of an {\it emergent time} and we can label the eigenvalues in ascending order of magnitude.  Key features of these eigenvalues are that they are symmetric about $0$, that the maximal eigenvalue scales as $\sqrt{N}$ when $N \rightarrow \infty$ \cite{Nish1}, and that hence the separation between eigenvalues scales as $\sqrt{N}^{-1}$. Hence, in the $N \rightarrow \infty$ limit, a continuous time emerges which runs from $- \infty$ to $+ \infty$.

The spatial matrices $A^i$ ($i = 1, \cdots, 9$) are not diagonal. Following \cite{Nish2},  we can define submatrices ${\tilde{A}}^i(n_i, t)$ by considering $n_i \times n_i$ matrices with centers a distance $t$ along the diagonal (see Fig. 3).

\begin{figure}
\includegraphics[scale = 0.7]{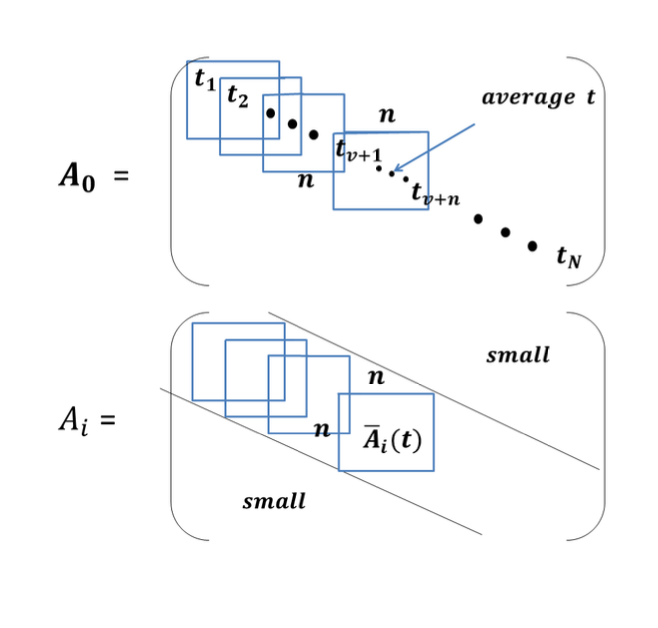}
\caption{Temporal matrix $A \equiv A_0$ (top panel) and spatial matrices $A_i$ (bottom panel)  in the basis in which the temperal matrix is diagonal. The spatial matrices $A_i$ (bottom panel) have ``block-diagonal form'' and can be used to define the sizes of the spatial dimensions at time $t$ via sub-matrices ${\bar{A_i}}(n_i, t)$ of $A_i$ centered a ``distance'' $t$ along the diagonal of $A_i$.  The size of the $n_i \times n_i$ (in the figure the subscript $i$ on $n_i$ is omitted) submatrices can vary between $1$ and $\sqrt{N}$. This figure is taken from \cite{Ydri} with permission.} 
\label{matrix}
\end{figure}

There has been a lot of interest in the time evolution of the {\it extent of space parameters}
\be \label{extent}
R_i(t, n_i)^2 \, \equiv \, \left\langle \frac{1}{n_i} \text{Tr} ({\bar{A_i}})(n_i, t)^2 \right\rangle \, ,
\ee
where the pointed angles indicate quantum expectation values. There are numerical indications \cite{IKKTpt} (see e.g. \cite{JNrev} for reviews) that there is a $SO(9) \rightarrow SO(3) \times SO(6)$ spontaneous symmetry breaking which takes place: precisely three of the extent of space parameters $R_i$ grow in time while the other six remain time independent \footnote{The numerical analysis is very involved. For the Euclidean matrix model, it is necessary to introduce cutoffs to make render the path integral well defined, and for the Lorentzian theory one is faced with rapidly oscillating exponents which are not easy to deal with. Hence, there is not yet universal agreement on the results (see e.g. \cite{IKKTnew} for recent discussions).}.  This result is supported by computations using the Gaussian expansion method,  where it can be shown \cite{Nish3} that the state with $SO(3) \times SO(6)$ symmetry has a lower free energy than the state with $SO(9)$ symmetry.  Fermions play an important role in deriving this result. In the absence of fermions, the symmetric state would be preferred \footnote{It can also be shown that the state with $SO(9)$ symmetry is not the one which minimizes the free energy \cite{BBL4}. The presence of fermions is again crucial to obtain this result.}.

In \cite{SB2}, the $n_i$ were identified with comoving spatial coordinates, and the extent of space parameter associated with $n_i$ was interpreted as the length of a curve along the i'th coordinate axis from coordinate value $0$ to $n_i$. Given this interpretation, we can define the spatial metric components $g_{ii}(t)$ (and using the remaining $SO(3)$ symmetry the full spatial metric tensor $g_{ij}(t)$). It was found that for $n_i \leq n_c \sim \sqrt{N}$ the emergent metric is spatially flat.

In the following section we will represent the expectation value of the matrices $A_{\mu}$ by a classical matrix$A_{\mu}^q$ defined by
\be
(A_{\mu}^q)^2 \, \equiv \, \left\langle  (A_{\mu})^2 \right\rangle \, .
\ee
In the classical limit (e.g. when the extent of space parameters are large) these matrices are a good representation of the quantum state.

\section{D-Strings in the Background Space-Time}

Classical solutions of the IKKT matrix model equations which represent an infinitely long D1 string about a vacuum were constructed in \cite{IKKT}.  The ansatz for a string in the 1-direction is
\bea
A_0 \, &=& \, \frac{T}{\sqrt{2 \pi n}} q \, \nonumber \\
A_1 \, &=& \, \frac{L}{\sqrt{2 \pi n}} p \, \\
A_i \, &=& \, 0 \,\,\, ( i = 2, \cdots, 9) \, , \nonumber
\eea
where $q$ and $p$ are $n \times n$ Hermitean matrices which satisfy the  commutation relations
\be
[q, p] \, = \, i \, ,
\ee
and have the eigenvalue distribution
\bea
0  &\leq&  q  \leq  \sqrt{2 \pi n} \, \nonumber \\
0  &\leq& p \leq  \sqrt{2 \pi n} \, .
\eea
 Note that $T$ and $L$ can be viewed as compactification radii in the temporal and spatial directions, respectively. This D1 string can be viewed as being made up of a string of instantons whose space-time positions are given by the eigenvalues of $q$ and $p$ \cite{Tseytlin}. Note that the fermionic matrices vanish.
 
Based on the abovementioned interpretation of the D1 string as an ensemble of constituent instantons, the ansatz for a pair of parallel D1 strings separated in the 2 direction by a distance $b$ is given by combining two blocks corresponding to individual D1 strings into a larger matrix, with vanishing off-diagonal matrix entries between the blocks:
\be \label{parallel0}
A_0 \, = \, 
\begin{bmatrix}
\frac{T}{\sqrt{2 \pi n}} q & 0 \\
0 & \frac{T}{\sqrt{2 \pi n}} q'
\end{bmatrix}
\ee
\be \label{parallel1}
A_1 \, = \, 
\begin{bmatrix}
\frac{L}{\sqrt{2 \pi n}} p & 0 \\
0 & \frac{L}{\sqrt{2 \pi n}} p'
\end{bmatrix}
\ee
\be \label{parallel2}
A_2 \, = \, 
\begin{bmatrix}
b/2 & 0 \\
0 & - b/2
\end{bmatrix}
\ee
with the other $A_i$ vanishing.  For two identical branes we have $q' = q$ and like $p' = p$.  

The force between parallel D1 strings can then be determined \cite{IKKT} by computing the one loop effective action obtained by expanding the full supersymmetric matrix action about this background.  This computation is done by inserting the ansatz
\bea
A_{a} \, &=& \, A_{a}^b + \delta A_{a} \nonumber \\
\psi_{\beta} \, &=& \, \delta \psi_{\beta} 
\eea
(where $A_a^b$ represent the D1 string background we are considering, and $\delta A_a$ and $\delta \psi$ are fluctuations) into the IKKT action and performing a one loop functional integral calculation to compute the effects of the matrix fluctuations (the fermionic matrices have trivial background) on the background.  It is found that in the supersymmetric matrix model there is no force between two parallel D1 strings \cite{IKKT}.

 In this paper we are interested in D1 string excitations not about the vacuum, but about the state of the matrix model given by the quantum mechanical partition function, the state which is seen to experience spontaneous breaking of the $SO(9)$ symmetry. This state is given by matrix ensembles $A_a \equiv A_a^q$ (where the superscript $q$ stands for quantum background). We propose the following ansatz for a pair of parallel D1 strings on top of the quantum background state $A_{\mu}^q$:
 \be
 A_{\mu}^b \, = \, A_{\mu}^q + A_{\mu}^{D1} \, ,
 \ee
 where $A_{\mu}^{D1}$ are the matrices corresponding to the two parallel D1 strings given above in (\ref{parallel0}, \ref{parallel1} ,  \ref{parallel2}), written in the basis in which the temporal matrix $q$ is diagonal.
  
In the following we will determine the one loop effective potential between the two branes. Our starting point is the expression derived in the appendix of \cite{IKKT} for the one loop effective action $W$ about a particular background configuration $p_{\mu}$ of the matrices $A_{\mu}$, i.e.
\be
A_{\mu} \, = \, p_{\mu} + a_{\mu} \, ,
\ee
where $a_{\mu}$ are the fluctuations.  The one loop action is
\bea \label{Weff}
W \, &=& \, \frac{1}{2}  {\tilde{\rm{Tr}}} {\rm{log}} \bigl( P_{\lambda}^2 \delta_{\mu \nu} - 2i F_{\mu \nu} \bigr) \nonumber \\
&& - \frac{1}{4} {\tilde{\rm{Tr}}} {\rm{log}} \bigl(  ( P_{\lambda}^2 + \frac{i}{2} F_{\mu \nu} \Gamma^{\mu \nu} \bigr) ( \frac{1 + \Gamma_{11}}{2} )  \nonumber \\
&& - {\rm{Tr}} {\rm{log}} ( P_{\lambda}^2 )
\eea
where ${\tilde{\rm{Tr}}}$ is a trace over both the vector / spinor indices and a matrix trace,  and ${\rm{Tr}}$ denotes the simple matrix model trace. In the above we have used the notation
\be
[p_{\mu}, X] \, \equiv \, P_{\mu}X 
\ee
and
\be
[f_{\mu \nu}, X] \, = \, F_{\mu \nu}X 
\ee
with
\be
f_{\mu \nu} \, = \, i [p_{\mu}, p_{\nu}] \, .
\ee
The first line in (\ref{Weff}) represents the effect of bosonic matrix fluctuations, the second line the effects of fermionic fluctuations, and the final line is the effect of the ghosts which are introduced because of gauge fixing. Note that $(1 + \Gamma_{11})/2$ is the analog of the four dimensional projection matrix $(1 + \Gamma_5)/2)$, and the $\Gamma^{\mu \nu}$ is the symmetric matrix constructed from the two-dimensional Dirac matrices in the standard way (see e.g. \cite{susy}).

In \cite{IKKT} it was shown that the one loop effective action between two parallel D strings of the form given in (\ref{parallel0}) vanishes. Supersymmetry is key to obtaining this results since it is only with supersymmetry that the contributions of the bosonic and fermionic degrees of freedom to the one loop effective potential cancel.  In fact, each of the three lines in the expression (\ref{Weff}) becomes a certain coefficient multiplying ${\rm{Tr}} P_{\lambda}^2$, and in the case of supersymmetry the coefficients add up to zero (see below).
 
 What we will show in the following is that the one loop effective action $W$ for the ansatz (\ref{exp}) representing parallel strings on the cosmological background factorizes into a term which depends on the background and a second term which has the same form as in the case of parallel D1 strings in the absence of a background, and hence does not depend on the distance $b$ between the strings. Thus, the conclusion that the force between parallel strings vanishes carries over.

To see this, we start with our ansatz
\be \label{exp}
p_{\mu} \, = \, A_{\mu}^q + A_{\mu}^{D1} \, ,
\ee
and in the following we will consider the individual terms which arise in the effective action. 

Let us first consider $F_{\mu \nu}$.  We insert the expansion (\ref{exp}) of the matrices into the definition of $F_{\mu \nu}$. Note  that $F_{\mu \nu}$ is a sum of a term which depends only on $A_{\mu}^q$ and gives the background value $F_{\mu \nu}^q$,  terms linear in $A_{\mu}^{D1}$, and a term which is quadratic in $A_{\mu}^{D1}$. The final term vanishes as it did in the analysis of \cite{IKKT} since the commutator of two $A_{\mu}^{D1}$ matrices are multiples of the identity matrix, and hence the corresponding operator vanishes. The terms linear in $A_{\mu}^{D1}$ are antisymmetric in $\mu, \nu$ and hence vanish when taking the trace (as in the first line of (\ref{Weff})) or when contracting with the symmetric matrix $\Gamma^{\mu \nu}$ (as in the second line of (\ref{Weff})). Thus, effectively
\be \label{res1}
F_{\mu \nu} \, = \, F_{\mu \nu}^q \, .
\ee
The vanishing of the linear terms is in this context the nontrivial realization.

Next,  consider the operator $P_{\lambda}^2$. Inserting the ansatz (\ref{exp}) there are terms independent of $A_{\mu}^{D1}$, terms linear in $A_{\mu}^{D1}$ and quadratic terms (which are independent of $A^q$).  Due to the cyclicity of the trace it can be seen that after taking the trace, the linear terms in $A_{\mu}^{D1}$ cancel (see the Appendix). Thus
\be \label{res2}
P_{\lambda}^2 \, = \, (P_{\lambda}^q )^2 + (P_{\lambda}^{D1})^2 \, ,
\ee
where $P_{\lambda}^{D1}$ is the $P$ operator constructed from the $A_{\mu}^{D1}$ matrices.

Combining the results (\ref{res1}) and (\ref{res2}) it follows that
\bea
W \, &=& \, W^q \nonumber \\
&+& \frac{1}{2}  {\tilde{\rm{Tr}}} {\rm{log}} \delta P_{\lambda}^2 \delta_{\mu \nu} - \frac{1}{4} {\tilde{\rm{Tr}}} {\rm{log}} \bigl(  ( \delta P_{\lambda}^2 ) ( \frac{1 + \Gamma_{11}}{2} ) 
\nonumber \\ & & - {\rm{Tr}} {\rm{log}} ( \delta P_{\lambda}^2 ) \\
&=& \, W^q \nonumber 
 \eea
since the terms in the second and third line are the corresponding terms for the IKKT ansatz \cite{IKKT} for two parallel D1 strings which add up to zero (see below in (\ref{cancel})). Hence, we see that the one loop effective action is independent of the string separation $b$, and that hence the force between parallel D1 strings on our cosmological background vanishes (at least computed to leading order).

We could also consider another ansatz for a set of two parallel D1 strings on our cosmological background, namely
 \be
A_0 \, = \, 
\begin{bmatrix} A_0^q & 0 & 0 \\
0 & \frac{T}{\sqrt{2 \pi n}} q & 0 \\
0 & 0 & \frac{T}{\sqrt{2 \pi n}} q'
\end{bmatrix}
\ee
\be
A_1 \, = \, 
\begin{bmatrix}
A_1^q & 0 & 0 \\
0 & \frac{L}{\sqrt{2 \pi n}} p & 0 \\
0 & 0 & \frac{L}{\sqrt{2 \pi n}} p'
\end{bmatrix}
\ee
\be
A_2 \, = \, 
\begin{bmatrix}
A_2^q & 0 & 0 \\
0 & b/2 & 0 \\
0 & 0 & - b/2
\end{bmatrix}
\ee
\be
A_a \, = \, 
\begin{bmatrix}
A_a^q & 0 & 0 \\
0 & 0 & 0 \\
0 & 0 & 0
\end{bmatrix}
\ee
for $a = 3, \cdots, 9$. Here, the $A_a^q$ are Hermitean $(N - 2n) \times (N - 2n)$ matrices, rendering the total matrices to have dimension $N$. We consider the limit $N \rightarrow \infty, n \rightarrow \infty$ with $n \ll N$.

For notational simplicity we will combine the $2n \times 2n$ blocks representing the two parallel branes by single blocks $A_a^{1,2}$. With this notation, the above matrices become
 \be \label{newansatz}
A_a\, = \, 
\begin{bmatrix} A_a^q & 0 \\
0 &  A_a^{1,2}
\end{bmatrix}
\ee
for $a = 0, \cdots, 9$.

Note that the commutators retain the block-diagonal form of the matrices. We make use of the fact that for $F_{\mu \nu} = 0$ for the block representing the two D-strings. In this case, the first term in (\ref{Weff}) (which we call $W1$) becomes
\be
W1 \, = \frac{1}{2} {\tilde{\rm{Tr}}} 
\begin{bmatrix}
{\rm{log}} \bigl( (P_{\lambda}^q)^2 \delta_{\mu \nu} - 2i F_{\mu \nu}^q \bigr) & 0 \\
0 & {\rm{log}} ( P_{\lambda}^{1,2})^2 \delta_{\mu \nu} 
\end{bmatrix} \, .
\ee
Since the trace is additive we get
\bea
W1 \, &=& \, \frac{1}{2} {\tilde{\rm{Tr}}} {\rm{log}} \bigl( (P_{\lambda}^q)^2 \delta_{\mu \nu} - 2i F_{\mu \nu}^q \bigr) \nonumber \\
&& + \frac{1}{2} {\tilde{\rm{Tr}}} {\rm{log}} ( P_{\lambda}^{1,2})^2 \delta_{\mu \nu} \, .
\eea
The first term is independent of the separation between the two D-strings, and the second term is identical to the corresponding one in the case of the effective potential of two parallel strings in the absence of the background. Similarly, the second term of the effective action (denoted by $W2$) separates
\bea
W2 \, &=& \, - \frac{1}{4} {\tilde{\rm{Tr}}} {\rm{log}}
\bigl(  ( (P_{\lambda}^q)^2 + \frac{i}{2} F_{\mu \nu}^q  \bigr) \Gamma^{\mu \nu} ( \frac{1 + \Gamma_{11}}{2} )  \nonumber \\
&& - \frac{1}{4} {\tilde{\rm{Tr}}} {\rm{log}}
\bigl(  ( (P_{\lambda}^{1,2})^2 ( \frac{1 + \Gamma_{11}}{2} ) \, .
\eea
Again, the first line depends only on the background, and the second line is the same result as the fermionic contribution in the case of two parallel D-strings in the absence of the background. Finally, the third term in (\ref{Weff}) (called $W3$) is
\be
W3 \, = \, - {\rm{Tr}} {\rm{log}} (P_{\lambda}^q)^2 - {\rm{Tr}} {\rm{log}} (P_{\lambda}^{1,2})^2 \, ,
\ee
where the first term only depends on the background, and the second one is the same as in the case of two parallel strings with no background.

Thus, we find for the one loop effective action
\be
W \, = \, W^q + W^{1,2} 
\ee
where $W^q$ depends only on the background, and the entire dependence on the separation between the two strings is contained in $W^{1,2}$, and this quantity is the same as in the case studied in \cite{IKKT}. Specifically, performing the vector and spinor traces yields
\be \label{cancel}
W^{1,2} \, = \, \bigl( \frac{10}{2} - \frac{16}{4} - 1 \bigr) {\rm{Tr }} {\rm{log}} ( P_{\lambda}^{1,2} )^2 \, = \, 0 \, .
\ee
We thus reach the same conclusion that in our cosmological background, there is no force between two parallel D-strings. Note that supersymmetry of the action is important to obtain this result, but supersymmetry of the background is not (our background breaks supersymmetry).

\section{Dynamical Decompactification Scenario}
 
We now argue that the results of the previous section provide a physical understanding of the $SO(9) \rightarrow SO(3) \times SO(6)$ symmetry breaking which allows precisely three of the nine extent of space parameters to become large (and hence classical).  We propose a realization of the suggestions of \cite{BV} in the context of the matrix model.

The nontrivial state of the IKKT matrix model which we are considering results from a thermal state of the BFSS model, as explained in \cite{SB1, SB2}. In the same way that a thermal state in a particle physics model contains all particle excitations which the model admits, distributed thermally, and with vanishing total charges, our state of the  BFSS model contains excitations of all string and brane degrees of freedom with vanishing total charge. In particular, it contains excitations of D1 strings oriented in all directions.  These strings have infinite length in comoving coordinates in the limit $N \rightarrow \infty$. Since the energy of a D1 string grows as the physical length increases, a condition of the emergence of large spatial dimensions is that the D1 strings extending along those directions must disappear. Since the winding number of a single D1 string is a topological invariant,  the only way for the winding strings to disappear is for two strings with opposite winding to collide and annihilate.

Since in our supersymmetric matrix model there are no long range forces between parallel D1 strings, the annihilation of the strings cannot happen in more than three large spatial dimensions since the probability of two string world sheets intersecting vanishes in more than four space-time dimensions. This argument provides a physical understanding of the observed $SO(9) \rightarrow SO(3) \times SO(6)$ symmetry breaking.

The absence of long range forces between D1 strings is a key ingredient in the above argument. Hence,  our physical argument implies that in the absence of supersymmetry, there will be no symmetry breaking and there are no obstacles to all spatial dimensions becoming large. This result agrees with numerical studies of IKKT type matrix models \cite{JNrev}, and with Gaussian expansion calculations \cite{Nish3} which indicate that in a purely bosonic matrix model the state with $SO(9)$ symmetry has lower free energy than states with the symmetry broken to $SO(p) \times SO(9 - p)$ ($p = 1, \cdots 4$), a result which was also seen in the BFSS matrix model \cite{BBL4}.

The nontrivial state of the matrix model contains other excitations beyond the D1 strings, e.g. D3 branes.  Such excitations can annihilate in more than three spatial dimensions, thus leaving the D1 strings as the last ones to persist, and those which determine the number of spatial dimensions which are finally allowed to grow large (see e.g. \cite{Stephon}).

\section{Discussion}

In this paper we have proposed a physical understanding of the origin of the $SO(9) \rightarrow SO(3) \times SO(6)$ symmetry breaking which is observed in the IKKT matrix model and shows the emergence of a three-dimensional large classical space from a starting point which looks more like a fuzzy 9-sphere.  Our proposal is based on embedding the argument of \cite{BV} in String Gas Cosmology into the matrix model.

Starting from a nontrivial state of the matrix model which emerges from a high temperature thermal state of the BFSS matrix model we have argued that extent of space parameters can only become large if the D1 strings which are initially present in the state can annihilate. Since in a supersymmetric matrix model there is no long range force between parallel D1 strings, as we have argued here,  no more than three spatial dimensions can become large since in a larger number of spatial dimensions the string world sheets have vanishing intersection probability, thus preventing annihilation.

There are lots of open issues. Most importantly, we need a better justification for why the ansatz for a configuration of two parallel D1 strings on top of our nontrivial background is the right one.  It is also important to demonstrate that the D1 strings in the matrix model are stable and in particular cannot break.

Another issue which we are currently investigating is whether the conclusions obtained here for the IKKT matrix model can also explain the origin of symmetry breaking in the BFSS matrix model (which we have used to obtain an emergent metric space-time in \cite{SB1, SB2}).  In that proposal, the emergent space-time was given by the Matsubara zero modes of the bosonic matrices, whose action is the same as the bosonic part of the IKKT action.  Since we are starting with a thermal state of the BFSS matrix model, it is reasonable to expect that there will be D1 string excitations in the IKKT picture (these would be the duals of the M-theory M2 branes). The BFSS matrix model is supersymmetric, and this should also be reflected on the IKKT side. 

In closing, we would also like to draw the attention of the reader to the work of \cite{Vafa} which proposes an embedding of String Gas Cosmology into a model of topological gravity.

\section*{Acknowledgments}

The research at McGill is supported in part by funds from NSERC and from the Canada Research Chair program.  We wish to thank many colleagues, in particular Suddho Brahma, Matthias Gaberdiel,  Hikaru Kawai, Samuel Laliberte and Edward Mazenc for listening to our thoughts and giving feedback. Particular thanks for Samuel Laliberte for sharing many ideas.  One of us (RB) also wishes to thank the Institutes of Theoretical Physics and of Particle and Astrophysics of the ETH Zurich for hospitality while this work was being completed.
 
\section*{Appendix}

In this appendix we show that the terms linear in $A_{\mu}^{D1}$ cancel in the evaluation of the operator $P_{\lambda}^2$ in (\ref{res2}).

The operator $P_{\lambda}^2$ acting on a matrix $X$ is
\bea
P_{\mu}^2 X \, &=& \, [P_{\mu}^q + P_{\mu}^{D1}, [P_{\mu}^q + P_{\mu}^{D1}, X]] \\
&=& \, (P_{\mu}^q)^2 X + P_{\mu}^q P_{\mu}^{D1} X + P_{\mu}^{D1} P_{\mu}^q X
+ (P_{\mu}^{D1})^2 X \, . \nonumber
\eea
 We now show that the terms linear in $P_{\mu}^{D1}$ cancel when taking the trace.  We write
 \bea \label{detail}
 P_{\mu}^q P_{\mu}^{D1} X \, &=& \, p_{\mu}^q p_{\mu}^{D1} X - p_{\mu}^q X p_{\mu}^{D1} \\ & & \,  - p_{\mu}^{D1} X p_{\mu}^q + X p_{\mu}^{D1} p_{\mu}^q \nonumber \\
 P_{\mu}^{D1} P_{\mu}^q X \, &=& \, p_{\mu}^{D1} p_{\mu}^q X - p_{\mu}^{D1} X p_{\mu}^q \nonumber \\
 && \, - p_{\mu}^q X p_{\mu}^{D1} + X p_{\mu}^q p_{\mu}^{D1} \, . \nonumber 
 \eea
 where the $p_{\mu}$ are the matrices associated with the operators $P_{\mu}$. Using the cyclicity of the trace, we find that adding the two lines of (\ref{detail}) the terms cancel pairwise. For example, the first term in the first expression cancels against the second term in the second expression.  Thus, we have established the result (\ref{res2}).

\end{document}